\documentclass{emulateapj}
\def\lax {\ifmmode{_<\atop^{\sim}}\else{${_<\atop^{\sim}}$}\fi}  
\def\gax {\ifmmode{_>\atop^{\sim}}\else{${_>\atop^{\sim}}$}\fi}  
\def\gtorder{\mathrel{\raise.3ex\hbox{$>$}\mkern-14mu
             \lower0.6ex\hbox{$\sim$}}}

\begin{document}

\title{On the nature of QPO phase lags in black hole candidates.}

\author{Nikolai Shaposhnikov\altaffilmark{1,2}}

\altaffiltext{1}{CRESST/University of Maryland, Department of Astronomy, College Park MD, 20742, nikolai.v.shaposhnikov@nasa.gov}

\altaffiltext{2}{Goddard Space Flight Center, NASA, 
Astrophysics Science Division, Greenbelt MD 20771; lev@milkyway.gsfc.nasa.gov}


\begin{abstract}

Observations of quasi-periodic oscillations (QPOs) in X-ray binaries  
hold a key to understanding many aspects of these enigmatic systems.
Complex appearance of the Fourier phase lags related to QPOs
is  one of the most puzzling observational effects in accreting black holes.
In this Letter we show that QPO properties, including phase lags, can be explained 
in a framework of a simple scenario, where the oscillating media provides a 
feedback on the emerging spectrum. 
We demonstrate that the QPO waveform is
presented by the product of a perturbation and a time delayed response factors, where the 
response is energy dependent. 
The essential property of this effect is its non-linear and multiplicative nature. 
Our multiplicative reverberation model successfully describes the QPO components in energy 
dependent power spectra as well as the appearance of the phase lags between 
signal in different energy bands.  We apply our 
 model to QPOs  observed by {\it Rossi X-ray Timing Explorer}
in  BH candidate XTE J1550--564. 
We briefly discuss the implications of the observed energy dependence 
of the QPO reverberation times and amplitudes to the nature of
the power law spectral component and its variability.



\end{abstract}

\keywords{accretion, accretion disks---black hole physics---X-rays: binaries---stars: individual (XTE J1550-564)}

\section{Introduction}

Galactic X-ray Black Hole (BH) candidates  
exhibit spectacular
evolution of timing and spectral behavior  \citep{rm,bell05}.
Strong variability is usually observed in states, dominated by non-thermal emission (i.e. hard and intermediate states).
The variability properties strongly correlate with overall 
spectral state evolution \citep{kw08}.
A common property of the variability is nearly periodic flux modulations with frequency between 0.01 and 20 Hz
generally referred to as low frequency quasi-periodic oscillations (QPOs). 
The exact origin of  QPOs in BH systems is still under debate. In this Paper we present a new important 
evidence which allows us to constrain the QPO models 
and to advance in understanding of this phenomenon and as well as the overall BH phenomelogy.
 
QPOs are observed as one or more  relatively narrow peaks in Fourier  power density spectra (PDS). 
QPO signal is shown to be strongly coupled with the power law part of the energy 
spectrum \citep{sh11,sz06}. Probably the most intriguing aspect of the QPO phenomenology 
is the behavior of Fourier phase lags, i.e. differences in phases 
calculated in two different energy ranges. High time resolution 
data on brightest BH transients, provided by  {\it Rossi X-ray Timing Explorer} ({\it RXTE})
allowed to study this effect in detail.
Phase lags observed during the initial intermediate state of  1998 event observed in XTE J1550-564
have shown complex behavior \citep{cui00,rem02}. Particularly remarkable 
is the fact that the fundamental QPO and its first harmonic exhibit lags
of opposite signs. While the fundamental has shown the negative (or soft) lags,
the lag in the first harmonics was hard for most observations (conventionally we define  positive or hard lags 
when signal is delayed at harder energies and vise versa for negative or soft lags). Similar
behavior was observed during 1999 outburst from XTE J1859+226 \citep{cas04} and 
in persistent BH candidate GRS  1915+105 \citep{lin00}. The conclusion of 
these studies was that either the lags behavior is due to different physical
origin of QPO harmonics or some unknown physical mechanism 
lead to a particular QPO waveform dependence on energy, which
is well reproduced in many observations. 
However, no QPO model have met the challenge to account the observed 
phase lag behavior.


In this Letter we present new and extremely important paradigm
 which allows much better understanding of QPO behavior and put a stronger
 constrains on its physical origin. First, we present a QPO waveform parametrization
 which provide proper description of QPO properties both in term of Fourier amplitudes
 and phases. In fact, we develop the first analytical model for QPO phase lags which allows
 direct modeling of the phase lags seen in observations. Secondly, 
 based on the results of our model, we draw conclusions on the physical 
 nature of the observed oscillations and the oscillating media. We tentatively call
 our model the "QPO reverberation model". We do not introduce any new physical 
paradigm. Instead, acting on the premises of the standard truncated disk/hot inner corona
model, we identify a specific way, in which all major observed QPO properties 
are naturally produced. The main idea behind the "QPO reverberation" 
model is that the inner part of the accretion flow acts as a non-linear 
forced oscillator where both perturbation and response signals 
enter the observed signal in multiplicative manner. While the perturbation
is hydrodynamical in nature, the response factor reflects the 
spectral shape reaction to the changes in oscillating media properties
introduced by the initial perturbation.

The QPO reverberation model matches the data very well, allowing to convert a complicated
form of the phase lags into well understood physical delay time.
The response signal is always delayed with respect to the 
perturbation, as expected in  case of forced oscillations.
The main conclusion which we draw based on our analysis
is that multiplicative reverberation is a primary mechanism leading to the non-sinusoidal 
nature of the QPO waveform and defining the behavior of QPO phase lags.

In the next section we describe the model and our {\it RXTE} data analysis. 
In \S 3 we briefly discuss the implications of our results to different  QPO scenarios.
Conclusions follow in the \S 4.

\section{Reverberation effect in QPO}

\subsection{Model}

We parametrize the QPO signal by the following
waveform:
\begin{equation}
s(t) \propto  (1+a e^{-\lambda t} \cos \omega_0 t) (1+b e^{-\lambda t} \cos (w_0(t-t_d))).
\label{qpo_waveform1}
\end{equation}

We assume that an oscillation process in the system signified by QPO 
leads to a time delayed response of the spectral form of the outgoing radiation.
The motivation for this functional form of the signal is most easily understood In terms 
of the power law shape of the average energy spectrum $NE^{-\alpha}$. If we assume
that the normalization $N$ and the index $\alpha$ experience exponentially decaying
oscillations, where oscillations in index are delayed with respect to the oscillations in 
the normalization then, after expanding the power law in Taylor series keeping only the first term,
we arrive to the expression (\ref{qpo_waveform1}). The Fourier transform of the waveform and its 
products (i.e. power spectrum, phase lags) are presented in Appendix. 

\begin{figure}[ptbptbptb]
\includegraphics[scale=0.6,angle=0]{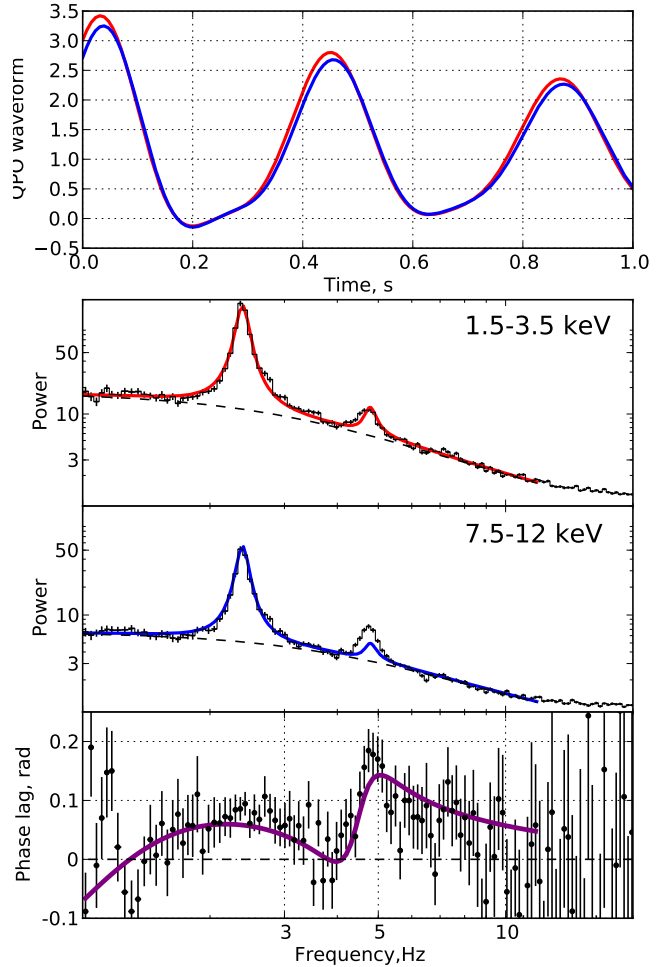}
\caption{{\it RXTE} data from observation of XTE J1550--564 (ObsID 30188-06-05-00). The bottom panel shows phase 
lags between signals in soft and hard  energy ranges. The middle panels show PDSs for the two energy ranges. The data are fitted with the reverberation model, shown as solid color lines (red for the soft range and blue for the  hard). For the PDS fits the dashed lines show contribution of Poisson  and broad-band noise components. The top panel shows the QPO waveforms for signal for both energies the soft (green) and the hard energy ranges as given by the model fit parameters.}
\label{1550_1}
\end{figure}

 \subsection{Application to data}
 
\begin{figure}[ptbptbptb]
\includegraphics[scale=0.6,angle=0]{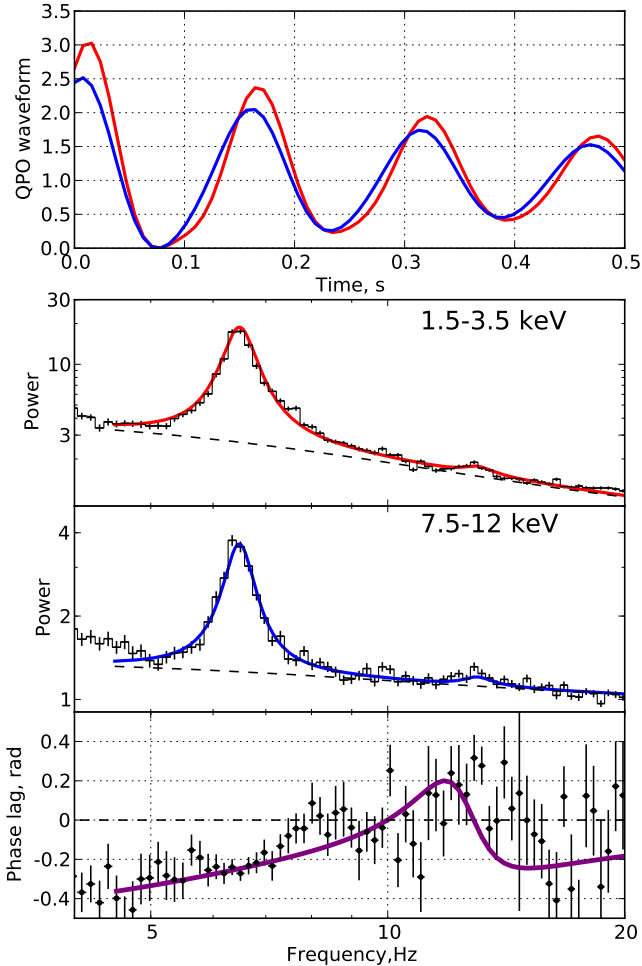}

\caption{ The same as Fig \ref{1550_1} for {\it RXTE} observation 30191-01-30-00 which is performed later 
during the outburst. Note that for this case, as opposed to the case presented in Fig \ref{1550_1}  the phase lags at the fundamental and the first harmonic have different  signs.
}
\label{1550_2}
\end{figure}

 We apply our model to the data collected with {\it RXTE} during the brightest 
 BH transient event observed from XTE J1550--564 \citep{hom01,cui00,st09}.
We use a combination of Binned and Event data modes to extract 
7.8125 millisecond time resolution lightcurves in channels covering
 1.5-3.5, 3.5-5.0,5.0-6.0, 6.0-7.5, 7.5-12.0, 12.0-14.6, 14.6-19.3 and 19.3 - 50.0 keV energy ranges. 
 We compute Fourier transforms for consecutive intervals 
of 64.0 sec for each lightcurve. We then compute corresponding power spectra and phase lags
as a complex argument of the cross spectrum using the first channel range as a reference.
We rebin these data products using 1.02 logarithmic rebinning factor.
 In Figures \ref{1550_1} and \ref{1550_2} we show representative power spectra and phase lags
 for two observations, 30188-06-05-00 (Observation 1) and 30191-01-30-00 (Observation 2).
  The fundamental QPO during these observations appeared at 2.35 Hz and at 6.5 Hz respectively. 
  The first observation is taken during a harder spectral state, i.e. earlier in the outburst. 
  It showed positive lags for both QPOs, while the second observation shows negative lag
 at the fundamental. For each observation we combine eight PDSs  and seven corresponding phase lags
 spectra in a joint model fit. Using the QPO waveform described by Eq. \ref{qpo_waveform1} 
 we model QPOs in the power spectra and corresponding phase lags between signals.
 The parameters describing perturbation signal $a$ and $\lambda$ are the same for all waveforms,
 while the parameters describing spectrum reverberation, i.e. its amplitude $b$
and time delay $t_d$ (or phase delay $\phi=2\pi \omega_0 t_d$) are individual for each energy range.
The data and the model fits for two energy ranges are presented in Fig. \ref{1550_1} and \ref{1550_2}.

\begin{figure}[ptbptbptb]
\includegraphics[scale=0.35,angle=-90]{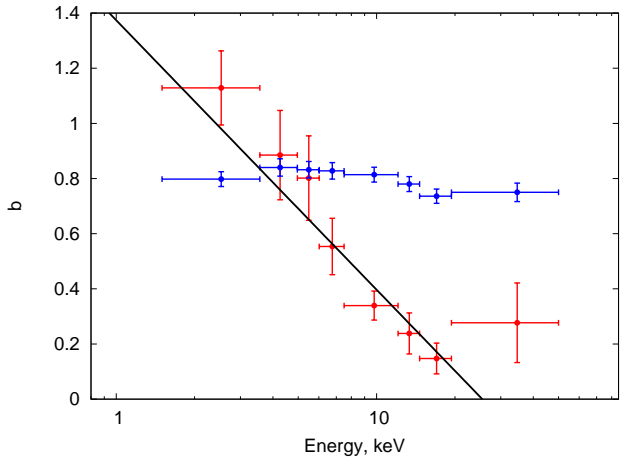}
\caption{ Energy dependence of the reverberation amplitude  $b$ for the Observation 1 (blue)
and Observation 2 (red).
The Observation 2 data is fitted with $f(E)=A\log(1/E) + B$ which is shown by black solid line.}
\label{bfig}
\end{figure}

Some systematic is seen in the phase lags fits in the vicinity of the
first harmonic of the QPO in power spectra. It would be naive to expect
that the parametrization using the expression (\ref{qpo_waveform1})
will describe every aspect of the data. Rather than the exact QPO waveform,
the presented model provides a basic framework for development of 
the future models which capture such effects as contribution of 
aperiodic PDS component in the phase lags, effects of statistical noise, etc.
Taking this fact, the model fits the
data extremely well capturing major behavior of the phase lags and the
exact form of the fundamental QPO, which is energy dependent and assymetrical.
In Figures \ref{tdelay} and \ref{bfig} we show the dependence of the reverberation times $t_d$
and amplitudes $b$ on energy.
Below we  discuss the implications of these results for the physical picture of QPO
production in BH candidates.



\section{Discussion \label{disc}}

The results of the presented model fits to the data show
that the reverberation scenario properly describes the QPO 
properties. Apparent non-linear multiplicative 
nature of the effect and its energy dependence naturally explains a
number of observational effects including shape of the QPOs in PDS,
energy dependence of the QPO powers, evolution and appearance 
of the phase lags. 

Probably even more important result of our analysis is the fact that
negative Fourier time lags observed in many cases for the fundamental QPO
are of a purely mathematical nature and {\it should not be treated as physical  times}.
The reverberation model properly  convolves a complicated behavior of Fourier  phases into
a one, well understood, physical time $t_d$ which can then be considered in terms of physical
models.

\begin{figure}[ptbptbptb]
\includegraphics[scale=0.35,angle=-90]{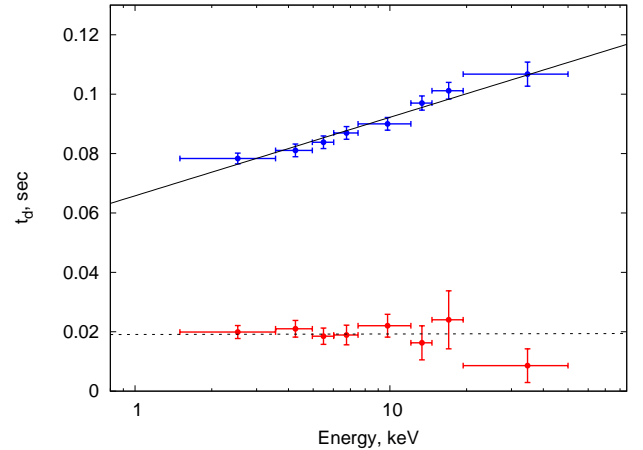}

\caption{ Energy dependence of the reverberation time  $t_d$ for the Observation 1 (blue points,
$a=1.30\pm 0.06$, $\lambda = 0.56\pm0.01$)
and Observation 2 (red points, $a=1.0\pm 0.1$, $\lambda = 2.07\pm0.03$).
The logarithm $f(E)=A\log(E) + B$ fit is shown by black solid line for the first observation and dashed line for 
the second one. The time delays for the first observations clearly follow the logarithm dependence while 
the data for the Observation 2 is consistent with the constant delays.}
\label{tdelay}
\end{figure}

Another important result is the nature of the observed reverberation and energy dependence of the
reverberation time and amplitude . It is clear that the perturbation actively changes the condition
in the system. Geometrical QPO models which are based on a mechanical
rotation or precession \citep[see e.g.]{sv98,sch06} lack this active ingredient. 
Reverberation effect strongly favors QPO models which involve oscillations affecting the state of 
matter which then is reflected in the form of the emerging spectrum. This group of models includes
 transition layer (corona) oscillations \citep{tlm98} and gravitationally trapped 
 disk oscillation modes \citep{wag01}. Logarithmic dependence of the time
 delays for the hard state (see FIg. \ref{tdelay} for Observation 1) strongly favors accretion induced 
 oscillations of the Compton corona as model for QPOs. 

In the truncated disk scenario a thin accretion disk extends
down to some inner radius where it superseded by a optically-thin geometrically thick configuration
or corona \citep[see e.g.][]{tom09}. 
In  this scenario, the thin disk is responsible for the soft thermal part of the spectrum,
while the corona produces a non-thermal power law part of the energy spectrum.
Variation in the inner disk radius, caused by surges of accreting matter, introduce
perturbation in the corona and modulate its physical parameters, i.e. its density,
temperature, optical depth, etc. This leads to temporal variations of the spectral
shape of the escaping radiation. The time delay between the perturbation and the
spectral response is set by sound speed in the corona. The logarithmic dependence of
time delays with energy observed for Observation 1 (hard state) indicates (see Figure \ref{tdelay}), according to
expectations, that Comptonization play significant part in spectral formation during these
states. Namely, multiple upscattering events experienced by photons 
cause wave-like structures to propagate from the low energies toward
the higher part of the time dependent spectrum. On the other hand, during the soft intermediate state (Observation 2), the reverberation time is flat with respect to energy, indicating that the spectrum reacting to the perturbation
at the same time at all energies. Moreover, the reverberation amplitude for this observation is well represented by
$b\propto \log(1/E)$, in agreement with variability being produced by a pure pivoting power law. 
This can not be facilitated through the thermal Comptonization
 and will require either synchrotron or non-thermal (such as bulk motion, see e.g. Titarchuk\&Shaposhnikov 2010) Comptonization mechanisms to be invoked. This subject is beyond this Letter and will be addressed in more detail in
upcoming publications.

\citet{mac11} studied QPOs observed in GRS 1915+105 by means of bispectrum and found
evidence for strong non-linear coupling between QPO harmonics and the broadband noise and considered
model involving non-linear oscillator.
Our findings also strongly suggest that the inner accretion flow region acts as a non-linear 
forced oscillator, which, by means of combination of radiation mechanisms such as thermal/non-thermal 
Comptonization and synchrotron radiation,  modulate the emerging spectrum of the system, leading to
a non-linear nature of the QPO signal. The waveform presented by expression (\ref{qpo_waveform1})
predicts that on time scales shorter than the exponential decay time $1/\lambda$ both the
average flux and variability are proportional to $e^{-\lambda t}$ and, therefore, are linearly 
related to each other. Rms-flux relation, well established in broad variability from BH sources, was recently
found in QPOs from XTE J1550-564 by \citet{heil11}. A specific feature of the relation
 is that it switches from positive to negative above 5 Hz.  \citet{heil11} show that
scaling of the QPO power to the power of underlying broad band noise corresponding to 
the QPO frequency reinstates the linear rms-flux relation, in excellent agreement with the
expectation that the noise variability is associated with the driving force for non-linear oscillator associated with 
QPO signal.

\section{Conclusions \label{summary}} 

We present a new observational effect in QPOs. We identify the time delayed reverberation 
in QPO signal as mechanism facilitating the phase lag behavior observed in bright galactic 
 BH candidates. The QPO reverberation model successfully 
 describes the QPOs in energy dependent power spectra and the corresponding phase lags.
 The presented results strongly point to oscillation of the Comptonizing and,most probably magnetized, 
 corona as the origin of the QPO signal from accreting BHs.
The QPO reverberation parametrization presents, for the first time,
a consistent description of the QPO signal both in Fourier amplitude and phase domains, providing an 
exciting avenue to finally resolving the QPO nature. 

\appendix

\section{ QPO waveform\label{waveform}}

Let us consider the waveform described by the expression:
\begin{equation}
s(t,E) \propto  (1+a e^{-\lambda t}\cos \omega_0 t) (1+b(E) e^{-\lambda t}\cos (\omega_0 t-\phi(E))),
\label{qpo_waveform}
\end{equation}
where $b(E)$ and $\phi(E)$ are functions of energy $E$.

The Fourier transform of $s(t)$ is given by
\begin{equation}
S(\omega,E)  =\frac{1}{2}\left[\frac{ab\cos \phi}{2\lambda+i\omega}+\frac{a+b e^{i\phi}}{\lambda+i(\omega-\omega_0)}+\frac{a+b e^{-i\phi}}{\lambda+i(\omega+\omega_0)} + \frac{ab e^{i\phi}}{2(2\lambda+i(\omega-2\omega_0))}+\frac{ab e^{-i\phi}}{2(2\lambda+i(\omega+2\omega_0))}\right]
\end{equation}

The QPO power spectrum and phase lags between signals at energies $E_1$ and $E_2$ are calculated in straightforward manner, i.e.:
\begin{equation}
P(E)=|S(E_{1,2})|^2
\end{equation}
and
\begin{equation}
\delta P(E_1,E_2) = \mathrm{Arg}(S(\omega,E_1)*\overline{S(\omega,E_2)})
\label{lag_model}
\end{equation}
correspondingly.


\begin{thebibliography}{}

\bibitem[Belloni(2005)]{bell05} Belloni, T.\ 2005, Interacting Binaries: Accretion, Evolution, and Outcomes, 797, 197 

\bibitem[Casella et al.(2004)]{cas04} Casella, P., Belloni, T., Homan, J., \& Stella, L.\ 2004, \aap, 426, 587 

\bibitem[Casella et al.(2005)]{cas05} Casella, P., Belloni, T., \& Stella, L.\ 2005, \apj, 629, 403 

\bibitem[Cui et al.(2000)]{cui00} Cui, W., Zhang, S.~N., \& Chen, W.\ 2000, \apjl, 531, L45 

\bibitem[Heil et al.(2011)]{heil11} Heil, L.~M., Vaughan, S., \& Uttley, P.\ 2011, \mnras, 411, L66 

\bibitem[Homan et al.(2001)]{hom01} Homan, J., Wijnands, R., van der Klis, M., et al.\ 2001, \apjs, 132, 377 

\bibitem[Klein-Wolt  \& van der Klis(2008)]{kw08} Klein-Wolt, M., \& van der Klis, M.\ 2008, \apj, 675, 1407 

\bibitem[Lin et al.(2000)]{lin00} Lin, D., Smith, I.~A., Liang, E.~P., B\"{o}ttcher, M.\ 2000, \apjl, 543, L141 

\bibitem[Maccarone et al.(2011)]{mac11} Maccarone, T.~J., Uttley, P., van der Klis, M., Wijnands, R.~A.~D., 
\& Coppi, P.~S.\ 2011, \mnras, 413, 1819 

\bibitem[McClintock et al.(2009)]{mr09} McClintock, J.~E., Remillard, R.~A., Rupen, M.~P., et al.\ 2009, \apj, 698, 1398 

\bibitem[Remillard et al.(2002)]{rem02} Remillard, R.~A., Sobczak, G.~J., Muno, M.~P., \& McClintock, J.~E.\ 2002, \apj, 564, 962 

\bibitem[Remillard \& McClintock(2006)]{rm} Remillard, R.~A., \& McClintock, J.~E.\ 2006, \araa, 44, 49 

\bibitem[Shaposhnikov \& Titarchuk(2009)]{st09} Shaposhnikov, N., \& Titarchuk, L.\ 2009, \apj, 699, 453 
 
 \bibitem[Shaposhnikov et al.(2011)]{sh11} Shaposhnikov, N., Swank, J.~H., Markwardt, C., \& Krimm, H.\ 2011, arXiv:1103.0531 
 
 \bibitem[Schnittman et al.(2006)]{sch06} Schnittman, J.~D., Homan, J., \& Miller, J.~M.\ 2006, \apj, 642, 420 

\bibitem[Sobolewska \&  \.{Z}ycki (2006)]{sz06} Sobolewska, M.~A., \.{Z}ycki, P.~T.\ 2006, \mnras, 370, 405 

 \bibitem[Stella \& Vietri(1998)]{sv98} Stella, L., \& Vietri, M.\ 1998, \apjl, 492, L59 

  \bibitem[Sunyaev \& Titarchuk (1980)]{st80} Sunyaev, R.A. \& Titarchuk, L.G. 1980,  A\&A, 86, 121 

\bibitem[Tomsick et al.(2009)]{tom09} Tomsick, J.~A., Yamaoka, K., Corbel, S., et al.\ 2009, \apjl, 707, L87 

\bibitem[Tagger \& Pellat(1999)]{tp99} Tagger, M., \& Pellat, R.\ 1999, \aap, 349, 1003 

\bibitem[Titarchuk et al.(1998)]{tlm98} Titarchuk, L., Lapidus, I., \& Muslimov, A.\ 1998, \apj, 499, 315 

\bibitem[Titarchuk \& Shaposhnikov(2010)]{ts10} Titarchuk, L., \& Shaposhnikov, N.\ 2010, \apj, 724, 1147 

\bibitem[Vignarca et al.(2003)]{vig03} Vignarca, F., Migliari, S., Belloni, T., Psaltis, D., \& van der Klis, M.\ 2003, \aap, 397, 729 

\bibitem[Wagoner et al.(2001)]{wag01} Wagoner, R.~V., Silbergleit, A.~S., \& Ortega-Rodr{\'{\i}}guez, M.\ 2001, \apjl, 559, L25 


\end{thebibliography}
\end{document}